%% file: ECOC_LaTeX_Template.tex
\documentclass[a4paper, oneside, twocolumn, notitlepage, 10pt, dvipsnames]{extarticle_ecoc}
\usepackage{ecoc}
\begin{document}
\selectlanguage{english}    % Standard Language

%-------------------------------------------------- Title -----------------------------------------------------%

\title{Lowering Error Floors for Hard Decision Decoding of OFEC Code}%

%------------------------------------------------- Authors-----------------------------------------------------%

\author{
    \vspace{-1mm}
    Jasper~Lagendijk,\textsuperscript{*} 
    Yunus~Can~Gültekin, 
    Alexios~Balatsoukas-Stimming,  
    Gabriele~Liga, 
    Alex~Alvarado 
    }

\maketitle                  % Create title and author

%------------------------------------------ Description of Authors ----------------------------------------------%

\begin{strip}
    \begin{author_descr}
        \vspace{-3mm}
        
        Department of Electrical Engineering, Eindhoven University of Technology,
        \textsuperscript{*}\textcolor{blue}{\uline{j.lagendijk@tue.nl}} 
        %\vspace{-3mm}
    \end{author_descr}
\end{strip}

% \setstretch{1.1}
%-------------------------------------------------- Footnote -------------------------------------------------------%
\renewcommand\footnotemark{}
\renewcommand\footnoterule{}
%\let\thefootnote\relax\footnotetext{text}

%-------------------------------------------------- Abstract ---------------------------------------------------------%
\begin{strip}
    \begin{ecoc_abstract}
        Stall patterns are known to cause an error floor in hard decision decoding of the OFEC code. We propose a novel stall pattern removal algorithm that lowers the error floor of state-of-the-art algorithms by an order of magnitude. \textcopyright2025 The Author(s)
    \end{ecoc_abstract}
    \vspace{-4mm}
\end{strip}

%-------------------------------------------------- Introduction Section -------------------------------------------------------%

\input{Sections/introduction}
\vspace{-2mm}
\input{Sections/ofec_structure}

\vspace{-2mm}
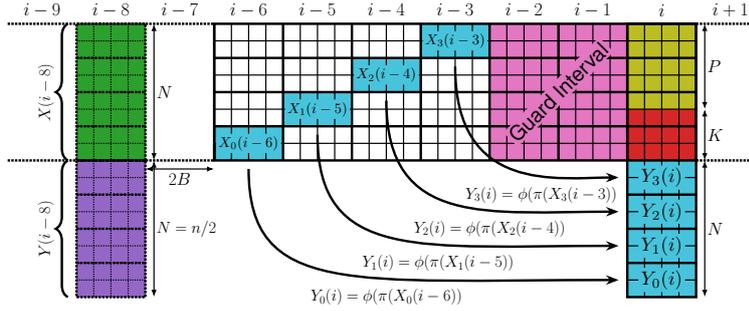
\begin{figure*}
    \centering
    \makebox[\textwidth][c]{% Force contents into a single line
        \begin{minipage}{0.64\textwidth}
            \begin{figure}[H]
                \centering
                \resizebox{\textwidth}{!}{\input{Figures/OFEC_encoding_medium.tikz}}
                \caption{Encoding of an OFEC code with $(n, k)=(16, 11)$.}
                \label{fig:OFEC_Encoding}
            \end{figure}
        \end{minipage}
        \hfill
        \begin{minipage}{0.35\textwidth}
            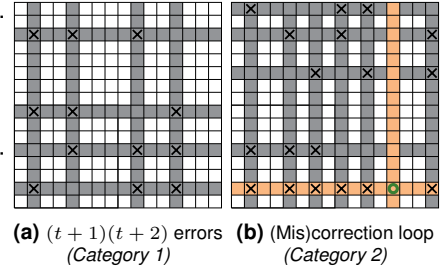
\begin{figure}[H]
                \centering
                \begin{subfigure}[b]{0.49\textwidth}
                    \centering
                    \resizebox{\textwidth}{!}{\input{Figures/PC_stall_patterns/PC_min_large}}
                    \caption{$(t+1)(t+2)$ errors \textit{(Category 1)}}
                    \label{fig:stall_PC_min_large}
                \end{subfigure}
                \hfill
                \begin{subfigure}[b]{0.49\textwidth}
                    \centering
                    \resizebox{\textwidth}{!}{\input{Figures/PC_stall_patterns/PC_large}}
                    \caption{(Mis)correction loop \textit{(Category 2)}}
                    \label{fig:stall_PC_large}
                \end{subfigure}
                \vspace{1em}
                \caption{Stall patterns in a PC ($t=2$).}
                \vspace{-.25em}
                \label{fig:PC_stall}
            \end{figure}
        \end{minipage}
    }
\end{figure*}

\input{Sections/stall_patterns}

\vspace{-2mm}
\input{Sections/stall_pattern_removal}

\vspace{-2mm}

\input{Sections/results}

\begin{figure}
    \centering
    \resizebox{\columnwidth}{!}{\input{Figures/SPR_pipeline.tikz}}
    \caption{Order and function of the SPR subroutines}
    \label{fig:spr_pipeline}
    \vspace{-2mm}
\end{figure}

\clearpage
\section{Acknowledgements}
The work of J. Lagendijk, Y. C. G{\"u}ltekin, A. Balatsoukas-Stimming and A. Alvarado is part of the project BIT-FREE with file number 20348 of the research programme Open Technology Programme which is (partly) financed by the Dutch Research Council (NWO).
%-------------------------------------------------- Bibliography Section -------------------------------------------------------%
% see also https://tex.stackexchange.com/questions/55030/text-before-references-but-after-bibliography-title-with-bibtex as of 2024-02-29
%\defbibnote{myprenote}{%
%Citations must be easy and quick to find. More precisely:
%\begin{itemize}
%    \item Please list all the authors. 
%    \item The title must be given in full length. 
%    \item Journal and conference names should not be abbreviated but rather given in full length.
%    \item The DOI number should be added incl. a link.
%\end{itemize}
%}
\printbibliography

\vspace{-4mm}

%%%%%%%%%%%%%%%%%%%%%%%%%%%%%%%%%%%%%%%%%%%%%

%---------------------------------------------- End of Document -----------------------------------------------%
\end{document}

%% file: Sections/introduction.tex
\section{Introduction}
\vspace{-2mm}
Modern high-speed communications rely heavily on optical networks.
%, where 
Data center interconnects and other short-distance links are currently attracting significant attention due to 
%artificial intelligence 
AI applications~\cite{barroso_datacenter_2018}. 
For these links, small form factor pluggable optical transceivers are used
%essential 
due to space constraints 
%within data centers
\cite{zhou_beyond_2020}. 
%The small form factor restricts heat dissipation which, together with the power constraints, creates a strict requirement on energy efficiency. 
Heat dissipation and power consumption create strict requirements on energy efficiency. 
%Satisfying this requirement while 
Achieving high throughput (800~Gbps and beyond) and reliability (post-FEC BER below $10^{-15}$) at low power usage demands well-designed DSP algorithms. 

A significant portion of the DSP power is used for the decoding of the FEC code~\cite{nagarajan_low_2021}. 
The complexity of the FEC decoder is therefore crucial to achieve the power and throughput targets.
The newest generation of pluggable optical transceivers, as defined in the 400G-800G OpenROADM and OpenZR+ standards~\cite{open_roadm_2022, open_zr_MSA}, uses the open FEC (OFEC) code. The OFEC code is a spatially coupled algebraic code using a $(256, 239)$ extended BCH (eBCH) code as component code~\cite{bose_class_1960, hocquenghem_codes_1959}. eBCH codes can be decoded using hard decision (HD) or soft decision (SD) decoding. The traditional algorithm for HD decoding of eBCH codes is bounded distance decoding (BDD). SD decoding is generally based on the Chase algorithm~\cite{chase_class_1972}. 

The advantage of the OFEC code, and by extension the class of spatially coupled codes, lies in the iterative nature of decoding. Iterative BDD (iBDD) is used when applying HD decoding and iterative Chase-Pyndiah is used for SD decoding~\cite{pyndiah_near-optimum_1998}. 
The state-of-the-art (SOTA) decoding of OFEC is hybrid, i.e., the component codes are decoded using a combination of SD and HD iterations~\cite{wang_real-time_2023}. This decoder, capable of achieving the required post-FEC BER of $10^{-15}$ at a pre-FEC BER of $0.02$~\cite{open_roadm_2022, open_zr_MSA}, uses 3 SD and 2 HD iterations~\cite{wang_real-time_2023}. 
The SD iterations use 93 Chase-Pyndiah test patterns, i.e., 93 BDD calls are made for each SD iteration compared to one BDD call per HD iteration. The SD iterations are about two orders of magnitude more complex than HD iterations.
Consequently, HD-only decoders achieving the target of $10^{-15}$ are of great interest.

Two issues arise when applying HD-only decoding to the OFEC code. 
Firstly, such decoders suffer from significant performance degradation compared to SD and hybrid decoders. 
Secondly, HD-only decoding of OFEC has been shown in~\cite{rapp_optimized_2024} to suffer from an error floor around $10^{-9}$ BER. 
This error floor is caused by error configurations that cannot be removed by applying iBDD, i.e., stall patterns.
In \cite{rapp_optimized_2024}, a stall pattern removal (SPR) algorithm was proposed.
%to remove this error floor. 
%This decoder consists of two parts: iBDD followed by an SPR algorithm. 
This SPR algorithm was shown to reduce the error floor below a BER of $10^{-10}$ for \emph{binary phase-shift keying}. 

In this work, we first show that the SPR algorithm in \cite{rapp_optimized_2024} also works for \emph{16-quadrature amplitude modulation} (16-QAM), which is the modulation considered in the standards~\cite{open_zr_MSA, open_roadm_2022}. However, we then show that this SPR algorithm suffers from a new error floor around $10^{-12}$ BER for 16-QAM.
We then propose a novel four-stage SPR algorithm, capable of lowering this new error floor by an order of magnitude, to a BER of $10^{-13}$. Through extrapolation, we estimate that our algorithm achieves a $0.15$~dB gain compared to the algorithm in~\cite{rapp_optimized_2024}.

%This novel algorithm consists of a total of 5-8 stages, including iBDD and four different SPR algorithms. The most aggressive combination of algorithms is shown to allow for stall pattern free decoding down to $10^{-13}$. Assuming no stall patterns exist causing error floors between $10^{-13}$ and $10^{-15}$, a delta net coding gain (NCG) of $0.25$~dB is expected. 

%% file: Sections/ofec_structure.tex
\section{Structure of the OFEC Code}

The OFEC code is a spatially-coupled code with an $(n, k)$ component code. 
%Here $n$ provides the block length of the component code and $k$ the number of information bits. 
An OFEC codeword is a semi-infinite matrix with an infinite number of columns and $N=n/2$ rows~\cite{zokaei_memory_2022}. %, when considering an $(n, k)$ component code. 
The OFEC code uses the $(256, 239)$ eBCH code as the component code\cite{open_roadm_2022,open_zr_MSA}. 
Fig.~\ref{fig:OFEC_Encoding} shows an example using the $(16,11)$ eBCH code as the component code for simplicity.
%Given an $(n, k)$ component code, the OFEC codeword is a matrix with $N = n/2$ rows (upper half of Fig.~\ref{fig:OFEC_Encoding}). 
Each column is a half eBCH codeword and $2B$ of these columns form a so-called \emph{chunk} (in the figure $B=2$). 
The bits highlighted in green in Fig.~\ref{fig:OFEC_Encoding} shows the chunk $X(i-8)$ at time instant $i-8$. 
Chunks are subdivided into $N/B=4$ blocks of $B\times2B$ bits. These blocks are labeled $X_j(i)$ with $0 \leq j < N/B$, where $X_0(i)$ is the lowest block in chunk $i$. Fig. \ref{fig:OFEC_Encoding} highlights the blocks $X_0(i-6)$ through $X_3(i-3)$ in blue, following a staircase pattern. 
Here, $X_0(i-6)$ is the lowest block of $X(i-6)$ and $X_3(i-3)$ the highest block of $X(i-3)$.   
The other half of each eBCH codeword consists of previously-encoded bits. These previously-encoded bits are stored in a buffer. 
During the encoding, these bits are obtained from the buffer and placed in a \emph{virtual chunk}. For chunk $X(i-8)$, these bits are placed in virtual chunk $Y(i-8)$, highlighted in purple in Fig.~\ref{fig:OFEC_Encoding}.
The previously-encoded bits are shared between the current eBCH codeword and a previously-encoded codeword, effectively coupling these codewords.
Each eBCH codeword in an OFEC codeword can be divided into three: $N$ previously-encoded (coupled) bits, $K$ new information bits, and $P=n-k$ parity bits. For time step $i$, these parts are highlighted in blue, red, and yellow in Fig.~\ref{fig:OFEC_Encoding}, resp.

Encoding chunk $X(i)$ requires obtaining the corresponding coupled bits from the buffer. This is done in a staircase pattern from previous chunks. A guard interval of two chunks is introduced, i.e., chunks $X(i-1)$ and $X(i-2)$ in Fig.~\ref{fig:OFEC_Encoding}.
To generate virtual chunk $Y(i)$, $N/B=4$ blocks are needed from the buffer. These blocks are labeled $Y_j(i)$ with $0 \leq j < N/B$. 
Here, $Y_0(i)$ is the lowest block in $Y(i)$. 
These blocks are obtained from chunks $X(i-N/B-2)$ through $X(i-3)$. Two interleaver functions, $\pi$ and $\phi$, are applied to each block individually and the resulting blocks are placed in virtual chunk $Y(i)$, as shown in Fig.~\ref{fig:OFEC_Encoding}. 
The virtual chunk $Y(i)$, combined with the information bits of chunk $X(i)$, creates the $2B$ information sequences necessary for the encoding of chunk $X(i)$.
The sequences are column-wise encoded, creating $2B$ eBCH codewords. The top $N$ bits of each column, which consist of the $K$ information bits and the $P$ parity bits, are transmitted. The bottom $N$ virtual bits are discarded. The encoding process can now start for chunk $i+1$. The OFEC code has parameters $B=16$ and $N=128$.

%% file: Figures/OFEC_encoding_medium.tikz
\definecolor{Cblue}{RGB}{31 119 180}
\definecolor{Cred}{RGB}{214 39 40}
\definecolor{Corange}{RGB}{255 127 14}
\definecolor{Cgreen}{RGB}{44 160 44}
\definecolor{Cpurple}{RGB}{148 103 198}
\definecolor{Cbrown}{RGB}{140 86 75}
\definecolor{Cpink}{RGB}{227 119 194}
\definecolor{Ccyan}{RGB}{23 190 207}
\definecolor{Cgray}{RGB}{127 127 127}
\definecolor{Cyellow}{RGB}{188 189 34}

\begin{tikzpicture}
    \foreach \b in {0, ..., 3}
    {
        \fill[color=SkyBlue] (0+4*\b, 2*\b) rectangle (4 + 4*\b, 2 + 2*\b);

        \fill[color=SkyBlue] (24, -8 + 2*\b) rectangle (28, -6 + 2*\b);
    }
    \fill[color=Cred] (24, 0) rectangle (28, 3);
    \fill[color=Cyellow] (24, 3) rectangle (28, 8);
    \fill[color=Cgreen] (-8, 0) rectangle (-4, 8);
    \fill[color=Cpurple] (-8, 0) rectangle (-4, -8);
    %\fill[color=Cbrown] (-16, 2) rectangle (-12, 4);
    \fill[color=Cpink] (16, 0) rectangle (24, 8);

    %% Basic OFEC matrix structure
    \draw[line width = 4pt] (0, 0) -- (0, 8) -- (28, 8) -- (28, 0) -- (0,0) -- (0, 1);

     \foreach \x in {1, ..., 27}
    {
        \draw[line width=2pt] (\x, 0) -- (\x, 8);
    }
    \foreach \y in {1, ..., 7}
    {
        \draw[line width=2pt] (0, \y) -- (28, \y);
    }

    \foreach \x in {0, 4, 8, 12, 16, 20, 24}
    {
        \draw[line width = 4pt] (\x, 0) -- (\x, 8);
    }

    \foreach \y in {2, 4, 6}
    {
        \draw[line width=4pt] (0, \y) -- (28, \y);
    }

    \foreach \t in {-2, 2, 6, 10, 14, 18, 22}
    {
        \pgfmathtruncatemacro\j{(26 - \t)*0.25}
        \node[] at (\t, 8.8) {\resizebox{!}{20pt}{\Huge $i - \j$}};
    }
    \node[] at (26, 8.8) {\resizebox{!}{20pt}{\Huge $i$}};

    %% Second chunk Y
    \foreach \x in {24, ..., 27}
    {
        \draw[line width=2pt] (\x, 0) -- (\x, -8);
    }
    
    \foreach \y in {-7, ..., -1}
    {
        \draw[line width=2pt] (24, \y) -- (28, \y);
    }

    \draw[line width=4pt] (24, 0) -- (24, -8) -- (28, -8) -- (28, 8) -- (24, 8) -- (24, 0);
    
    \foreach \y in {-6, -4, -2}
    {
        \draw[line width=4pt] (24, \y) -- (28, \y);
    }

    %% Lines between old and new chunks
    \draw[line width = 4pt, -Stealth] (2,-0.5) .. controls (2, -8) and (8, -7) .. (23.5, -7);

    \draw[line width = 4pt, -Stealth] (6, 1.5) .. controls (6, -6) and (12, -5) .. (23.5, -5);
    
    \draw[line width = 4pt, -Stealth] (10, 3.5) .. controls (10, -4) and (16, -3) .. (23.5, -3);
    
    \draw[line width = 4pt, -Stealth] (14, 5.5) .. controls (14, -2) and (20, -1) .. (23.5, -1);
    
    \node[] at (10, -8) {\resizebox{!}{20pt}{\huge $Y_0(i)=\phi(\pi(X_0(i-6))$}};
    \node[] at (13, -6) {\resizebox{!}{20pt}{\huge $Y_1(i)=\phi(\pi(X_1(i-5))$}};
    \node[] at (16, -4) {\resizebox{!}{20pt}{\huge $Y_2(i)=\phi(\pi(X_2(i-4))$}};
    \node[] at (19, -2) {\resizebox{!}{20pt}{\huge $Y_3(i)=\phi(\pi(X_3(i-3))$}};

    \draw[{Latex}-{Latex}, line width = 2pt] (28.5, 0) node[right, yshift=-4cm] {\resizebox{!}{20pt}{\huge $N$}} -- (28.5, -8);

    \draw[{Latex}-{Latex}, line width = 2pt] (28.5, 3) node[right, yshift=-1.5cm] {\resizebox{!}{20pt}{\huge $K$}} -- (28.5, 0);

    \draw[{Latex}-{Latex}, line width = 2pt] (28.5, 8) node[right, yshift=-2.5cm] {\resizebox{!}{20pt}{\huge $P$}} -- (28.5, 3);

    %\draw[{Latex}-{Latex}, line width = 2pt] (24, -8.5) node[below, xshift=2cm, yshift=-0.1cm] {\resizebox{!}{20pt}{\huge $2B$}} -- (28, -8.5);

    %%%%%%%%%%%%%%%%%%%%%%%%%%%%%%%%%%%%%%%%%%%%%%%%%%%%%%%%%%%%%%%%%%%%%%%%%%%%%%%%%%%%%%%%%%%%%%%%%%%
    %%%%%%%%%%%%%%%%%%%%%%%%%%%%%%%%%%%% Highlight chunk in i-8 %%%%%%%%%%%%%%%%%%%%%%%%%%%%%%%%%
    \draw[line width = 4pt, dotted] (-8, 0) -- (-4, 0) -- (-4, 8) --  (-8, 8) -- (-8, 0) -- (-4, 0);
    \draw[line width=4pt, dotted] (-8, 0) -- (-8, -8) -- (-4, -8) -- (-4, 0) -- (-8, 0);% -- (-4, -1);

    \draw [decorate,decoration={brace,amplitude=20pt,raise=0ex}, line width = 4pt]
  (-8.5,0) -- (-8.5,8) node[rotate=90, midway, yshift=0.5in]{\resizebox{!}{20pt}{\huge $X(i-8)$}};
      \draw [decorate,decoration={brace,amplitude=20pt,raise=0ex}, line width = 4pt]
  (-8.5,-8) -- (-8.5,0) node[rotate=90, midway, yshift=0.5in]{\resizebox{!}{20pt}{\huge $Y(i-8)$}};
    \foreach \x in {-7, ..., -5}
    {
        \draw[line width=2pt, dotted] (\x, 8) -- (\x, -8);
    }
    
    \foreach \y in {-7, ..., 7}
    {
        \draw[line width=2pt, dotted] (-8, \y) -- (-4, \y);
    }

    \foreach \y in {-6, -4, ..., 6}
    {
        \draw[line width=4pt, dotted] (-8, \y) -- (-4, \y);
    }

    \draw[line width=4pt, dotted] (-12, 0) -- (0, 0);
    \draw[line width=4pt, dotted] (-12, 8) -- (0, 8);
    \draw[line width=4pt, dotted] (28, 0) -- (32, 0);
    \draw[line width=4pt, dotted] (28, 8) -- (32, 8);

   % \foreach \x in {-16, -12, -8, 32}
    %{
    %    \draw[line width=2pt, dotted] (\x, 0) -- (\x, 8);
    %}

    \foreach \t in {-6, -10}
    {
        \pgfmathtruncatemacro\j{(26 - \t)*0.25}
        \node[] at (\t, 8.8) {\resizebox{!}{20pt}{\Huge $i - \j$}};
    }
    \foreach \t in {30}
    {
        \pgfmathtruncatemacro\j{(\t-26)*0.25}
        \node[] at (\t, 8.8) {\resizebox{!}{20pt}{\Huge $i + \j$}};
    }
    \draw[{Latex}-{Latex}, line width = 2pt] (-3.5, 0) node[right, yshift=-4cm] {\resizebox{!}{20pt}{\huge $N=n/2$}} -- (-3.5, -8);
    \draw[{Latex}-{Latex}, line width = 2pt] (-3.5, 8) node[right, yshift=-4cm] {\resizebox{!}{20pt}{\huge $N$}} -- (-3.5, 0);

    \draw[{Latex}-{Latex}, line width = 2pt] (-4, -0.5) node[below, xshift=2cm, yshift=-0.1cm] {\resizebox{!}{20pt}{\huge $2B$}} -- (0, -0.5);

    \node[fill=Cpink, anchor=center, rotate=45, rectangle, draw=Cpink] at (20, 4) {\resizebox{!}{25pt}{Guard Interval}};

    \node[anchor=center, rectangle, draw=SkyBlue, fill=SkyBlue] at (2, 1) {\resizebox{!}{18pt}{$X_0(i-6)$}};  
    \node[anchor=center, rectangle, draw=SkyBlue, fill=SkyBlue] at (6, 3) {\resizebox{!}{18pt}{$X_1(i-5)$}};  
    \node[anchor=center, rectangle, draw=SkyBlue, fill=SkyBlue] at (10, 5) {\resizebox{!}{18pt}{$X_2(i-4)$}};  
    \node[anchor=center, rectangle, draw=SkyBlue, fill=SkyBlue] at (14, 7) {\resizebox{!}{18pt}{$X_3(i-3)$}};  

    \node[anchor=center, rectangle, draw=SkyBlue, fill=SkyBlue] at (26, -7) {\resizebox{!}{24pt}{$Y_0(i)$}};      
    \node[anchor=center, rectangle, draw=SkyBlue, fill=SkyBlue] at (26, -5) {\resizebox{!}{24pt}{$Y_1(i)$}};      
    \node[anchor=center, rectangle, draw=SkyBlue, fill=SkyBlue] at (26, -3) {\resizebox{!}{24pt}{$Y_2(i)$}};      
    \node[anchor=center, rectangle, draw=SkyBlue, fill=SkyBlue] at (26, -1) {\resizebox{!}{24pt}{$Y_3(i)$}};

    %%%%%%%%%%%%%%%%%%%%%%%%%%%%% Highlight block in i-10 %%%%%%%%%%%%%%%%%%%%%%%%%%%%%%%%%%%
    
    %\draw [decorate,decoration={brace,amplitude=12pt,raise=0ex}, line width = 4pt] (-0.2,0.2) -- (-0.2,1.8) node[rotate=90, midway, yshift=0.5in, draw=white, fill=white]{\resizebox{!}{18pt}{\huge $X_0(i-6)$}};
   %\foreach \x in {-15, ..., -13}
    %{
    %    \draw[line width=2pt, dotted] (\x, 8) -- (\x, 0);
    %}
    
    %\foreach \y in {1, ..., 7}
    %{
    %    \draw[line width=2pt, dotted] (-16, \y) -- (-12, \y);
    %}

    %\foreach \y in {2, 4, ..., 6}
    %{
    %    \draw[line width=4pt, dotted] (-16, \y) -- (-12, \y);
    %}

    %\draw[{Latex}-{Latex}, line width = 2pt] (-11.5, 4) node[right, yshift=-1cm] {\resizebox{!}{20pt}{\huge $B$}} -- (-11.5, 2);
    %\draw[{Latex}-{Latex}, line width = 2pt] (-16, -0.5) node[below, xshift=2cm, yshift=-0.1cm] {\resizebox{!}{20pt}{\huge $2B$}} -- (-12, -0.5);

\end{tikzpicture}

%% file: Figures/PC_stall_patterns/PC_min_large.tex
\begin{tikzpicture}
   \tikzset{cross/.style={cross out, draw, 
         minimum size=2*(#1-\pgflinewidth), 
         inner sep=0pt, outer sep=0pt}}
    \draw[] (0, 0) -- (0, 8) -- (8, 8) -- (8, 0) -- (0, 0);

    \foreach \x/\y in {1/1, 9/4, 12/7, 4/13}
    {
        \fill[color=Gray] (\x, 0) rectangle (\x+1, 16);
        \fill[color=Gray] (0, \y) rectangle (16, \y+1);
    }

    \draw[] (0, 0) -- (0, 16) -- (16, 16) -- (16, 0) -- (0, 0);

    \foreach \x in {1, ..., 15}
    {
        \draw[] (0, \x) -- (16, \x);
        \draw[] (\x, 0) -- (\x, 16);
    }

    \draw(1.5, 1.5) node[cross=14pt, line width = 4pt]{};
    %\draw(1.5, 4.5) node[cross=14pt, line width = 4pt]{};
    \draw(1.5, 7.5) node[cross=14pt, line width = 4pt]{};
    \draw(1.5, 13.5) node[cross=14pt, line width = 4pt]{};

    %\draw(4.5, 1.5) node[cross=14pt, line width = 4pt]{};
    \draw(4.5, 4.5) node[cross=14pt, line width = 4pt]{};
    \draw(4.5, 7.5) node[cross=14pt, line width = 4pt]{};
    \draw(4.5, 13.5) node[cross=14pt, line width = 4pt]{};

    \draw(9.5, 1.5) node[cross=14pt, line width = 4pt]{};
    \draw(9.5, 4.5) node[cross=14pt, line width = 4pt]{};
    %\draw(9.5, 7.5) node[cross=14pt, line width = 4pt]{};
    \draw(9.5, 13.5) node[cross=14pt, line width = 4pt]{};

    \draw(12.5, 1.5) node[cross=14pt, line width = 4pt]{};
    \draw(12.5, 4.5) node[cross=14pt, line width = 4pt]{};
    \draw(12.5, 7.5) node[cross=14pt, line width = 4pt]{};
    %\draw(12.5, 13.5) node[cross=14pt, line width = 4pt]{};

\end{tikzpicture}

%% file: Figures/PC_stall_patterns/PC_large.tex
\begin{tikzpicture}
      \tikzset{cross/.style={cross out, draw, 
         minimum size=2*(#1-\pgflinewidth), 
         inner sep=0pt, outer sep=0pt}}
    \draw[] (0, 0) -- (0, 8) -- (8, 8) -- (8, 0) -- (0, 0);

    \foreach \x/\y in { 4/4, 6/4, 8/10, 10/10, 1/13, 15/15}
    {
        \fill[color=Gray] (\x, 0) rectangle (\x+1, 16);
        \fill[color=Gray] (0, \y) rectangle (16, \y+1);
    }
    \fill[color=Apricot] (0, 1) rectangle (16, 2);
    \fill[color=Apricot] (12, 0) rectangle (13, 16);

    \draw[] (0, 0) -- (0, 16) -- (16, 16) -- (16, 0) -- (0, 0);

    \foreach \x in {1, ..., 15}
    {
        \draw[] (0, \x) -- (16, \x);
        \draw[] (\x, 0) -- (\x, 16);
    }

    \draw(1.5, 1.5) node[cross=14pt, line width = 4pt]{};
    \draw(4.5, 1.5) node[cross=14pt, line width = 4pt]{};
    \draw(6.5, 1.5) node[cross=14pt, line width = 4pt]{};
    \draw(8.5, 1.5) node[cross=14pt, line width = 4pt]{};
    \draw(10.5, 1.5) node[cross=14pt, line width = 4pt]{};
    \filldraw[color=OliveGreen, fill=Apricot, line width = 7pt] (12.5, 1.5) circle (9pt);%node[cross=14pt, line width = 4pt, color=Maroon]{};
    \draw(15.5, 1.5) node[cross=14pt, line width = 4pt]{};

    \draw(1.5, 4.5) node[cross=14pt, line width = 4pt]{};
    \draw(4.5, 4.5) node[cross=14pt, line width = 4pt]{};
    \draw(6.5, 4.5) node[cross=14pt, line width = 4pt]{};
    %\draw(8.5, 4.5) node[cross=14pt, line width = 4pt, color=Maroon]{};
    %\draw(12.5, 4.5) node[cross=14pt, line width = 4pt]{};
    %\draw(15.5, 4.5) node[cross=14pt, line width = 4pt]{};

    \draw(10.5, 10.5) node[cross=14pt, line width = 4pt]{};
    %\draw(10.5, 4.5) node[cross=14pt, line width = 4pt]{};
    \draw(10.5, 15.5) node[cross=14pt, line width = 4pt]{};

    %\draw(1.5, 6.5) node[cross=14pt, line width = 4pt, color=Maroon]{};
    %\draw(4.5, 6.5) node[cross=14pt, line width = 4pt, color=Maroon]{};
    %\draw(6.5, 6.5) node[cross=14pt, line width = 4pt, color=Maroon]{};
    %\draw(8.5, 6.5) node[cross=14pt, line width = 4pt]{};
    %\draw(12.5, 6.5) node[cross=14pt, line width = 4pt]{};
    %\draw(15.5, 6.5) node[cross=14pt, line width = 4pt]{};

    %\draw(1.5, 10.5) node[cross=14pt, line width = 4pt, color=Maroon]{};
    %\draw(4.5, 10.5) node[cross=14pt, line width = 4pt, color=Maroon]{};
    \draw(6.5, 10.5) node[cross=14pt, line width = 4pt]{};
    %\draw(8.5, 10.5) node[cross=14pt, line width = 4pt]{};
    %\draw(12.5, 10.5) node[cross=14pt, line width = 4pt]{};
    \draw(15.5, 10.5) node[cross=14pt, line width = 4pt]{};

    %\draw(1.5, 13.5) node[cross=14pt, line width = 4pt, color=Maroon]{};
    \draw(4.5, 13.5) node[cross=14pt, line width = 4pt]{};
    %\draw(6.5, 13.5) node[cross=14pt, line width = 4pt]{};
    \draw(8.5, 13.5) node[cross=14pt, line width = 4pt]{};
    %\draw(12.5, 13.5) node[cross=14pt, line width = 4pt, color=Maroon]{};
    \draw(15.5, 13.5) node[cross=14pt, line width = 4pt]{};

    \draw(1.5, 15.5) node[cross=14pt, line width = 4pt]{};
    %\draw(4.5, 15.5) node[cross=14pt, line width = 4pt]{};
    %\draw(6.5, 15.5) node[cross=14pt, line width = 4pt]{};
    \draw(8.5, 15.5) node[cross=14pt, line width = 4pt]{};
    %\draw(12.5, 15.5) node[cross=14pt, line width = 4pt, color=Maroon]{};
    %\draw(15.5, 15.5) node[cross=14pt, line width = 4pt, color=Maroon]{};

\end{tikzpicture}

%% file: Sections/stall_patterns.tex
\section{Stall Patterns}
%When considering 
Stall patterns are of major concern for iBDD of product codes (PC), the OFEC code, and other spatially coupled codes. These patterns are configurations of errors, occurring in positions such that, regardless of the number of iBDD iterations, they cannot be removed. In the case of an eBCH code with error correcting capability $t$, these stall patterns can occur when each related codeword has more than $t$ errors. Two categories of stall patterns are considered in this paper, explained based on a PC for simplicity. 
The statements can be extended to the OFEC code straightforwardly. 
Fig.~\ref{fig:PC_stall} illustrates stall patterns from both categories.
%for a PC with $t=2$. 
%, with examples shown in Fig. \ref{fig:PC_stall}. 
%The first category is the minimal stall pattern, the pattern with the least amount of errors, as shown in Fig.~\ref{fig:stall_PC_min}. %, while still causing an error floor. 
%For PCs and OFEC, this stall pattern has size $(t+1)^2$. For this stall pattern, each of the $2(t+1)$ associated rows and columns will have $t+1$ errors and therefore fail to decode. 
%For the product code shown in Fig. \ref{fig:stall_PC_min}, each of the $t+1$ rows and each of the $t+1$ columns that are part of the stall pattern are marked in gray. Each error is marked by a cross. The second stall type pattern taken into consideration is the set of stall patterns with the next smallest number of errors. 

Category 1 consists of stall patterns in which each relevant eBCH codeword has $(t+1)$ errors, therefore iBDD fails. The smallest possible stall pattern for a PC, and by extension OFEC code, has $(t+1)^2$ errors and $2(t+1)$ failed rows and columns\cite{sukmadji_zipper_2022}. However, larger stall patterns also exist.
Category 1 is shown in Fig.~\ref{fig:stall_PC_min_large}, where each gray row or column has $(t+1)$ errors, and thus failed decoding.
Black crosses indicate errors.

Category 2 consists of stall patterns where codewords have up to $d_{\text{min}}$ errors, where $d_{\text{min}}$ is the minimum Hamming distance. A codeword with up to $d_{\text{min}}$ errors can still be decoded to a valid but incorrect codeword. Such a miscorrection can cause stall patterns with a correction/miscorrection loop~\cite{hager_approaching_2018}. %In such a loop, 
Given that for both PC and OFEC, a bit is protected by two codewords, a correction/miscorrection loop can occur if one of the codewords has at least $d_{\text{min}}-t$ errors, and the coupled codeword has less than $t$ errors. In such a case, when the first codeword is decoded, it is miscorrected, introducing up to $t$ new errors. Then, when the second codeword is decoded, one of these new errors is removed. This process will loop, miscorrected and corrected in each iteration. Fig.~\ref{fig:stall_PC_large} shows a stall pattern of Category 2. The orange row will lead to a miscorrection adding the error marked by the green circle. When decoding the orange column, this error is removed, creating the correction/miscorrection loop. 
 
%This category of stall patterns does not only contain rows/columns that fail to decode, but also rows/columns that are stuck in a correction/miscorrection loop. This correction/miscorrection loop consists of a row/column that has $t$ or less bits different than a valid codeword, different from the original transmitted codeword. The eBCH decoder will then decode this row/column to the valid but incorrect codeword, a miscorrection. During the decoding of the column/row, these newly added errors are corrected. The next iteration the error(s) will be added again.
%An example is shown in Fig. \ref{fig:stall_PC_large}. Here the error marked by a red cross will be miscorrected by each column decoding iteration and then corrected by each row decoding iteration. The two columns are then stuck in a loop, the correction/miscorrection loop. 
%The remaining stall patterns contain contain $(2t)^2$ or more errors. These stall patterns are not trivial to detect, as the eBCH decoder can no longer detect these errors. These stall patterns are however sufficiently large that they do not cause an error floor above $10^{-15}$ BER. 

%% file: Sections/stall_pattern_removal.tex
\section{Proposed Stall Pattern Removal Algorithm}
SPR algorithms aim to remove any residual errors caused by stall patterns. The algorithms are applied after the BDD decoding iterations. These algorithms consist of first detecting a stall pattern and then flipping bits at positions where errors are likely to have occurred. In the case of iBDD, the output of the BDD decoder can be used for this purpose. BDD can either correct errors, fail decoding, or determine that the input was already a valid codeword. 
These three output flags can be used to detect stall patterns and find the error positions. 

\begin{figure}[t]
    \centering
    %\vspace{-5mm}
    \hspace{-10mm}
    \resizebox{1\columnwidth}{!}{\input{Figures/OFEC_performance.tikz}}
    %\vspace{-5mm}    
    \caption{Performance of the discussed OFEC decoders.}% All decoders use 20 HD iBDD iterations}
    \label{fig:results}
    \vspace{-2.5mm}
\end{figure}

The SPR algorithm in \cite{rapp_optimized_2024} consists of flipping a bit if both associated codewords failed to decode. Applying this algorithm to the two stall patterns in Fig.~\ref{fig:stall_PC_min_large}, all bits that have both the row and column highlighted are flipped, removing all current errors. However, four new errors appear at the crossing points of the gray codewords without crosses. As there is now at most one error in a codeword, these errors can be resolved using additional BDD iterations after the SPR algorithm, i.e., cleanup BDD iterations. In the case of Fig. \ref{fig:stall_PC_large}, all bits that have both the row and column marked in gray will be flipped. The column marked in orange results in `correction with bits flipped', and thus none of its bits will be flipped by the SPR algorithm. After flipping the bits, most of the current errors are removed, but new errors are also added. These newly added bits will cause all columns to have $(t+1)$ errors. These will fail to decode, i.e., a new stall pattern is created. Fig.~\ref{fig:results} shows the performance of the algorithm proposed in~\cite{rapp_optimized_2024} for OFEC with 16-QAM. The initial error floor of the decoder without SPR at $10^{-9}$ is removed. However, our simulations uncovered a second error floor at $10^{-12}$ caused by larger stall patterns, both of category 1 and 2. Thus, more sophisticated SPR algorithms are needed to deal with these patterns. 

To remove this newly-found error floor, we propose a three-stage SPR algorithm (Fig.~\ref{fig:spr_pipeline}). Each stage of the SPR algorithm consists of one of three different subroutines, SPR1-3, each capable of dealing with a subset of the found stall patterns. SPR1 and SPR2 are novel, while SPR3 is a variation of the algorithm in \cite{rapp_optimized_2024}. 
Fig.~\ref{fig:spr_pipeline} shows a block diagram indicating the order of the stages. 
In addition, we use two versions of BDD: regular BDD, after each subroutine, and we propose the novel miscorrection reduction BDD (MRBDD), before each subroutine. MRBDD applies regular BDD, however, only the top $N$ bits of each codeword (see Fig.~\ref{fig:OFEC_Encoding}) are updated. The bits in the virtual chunk are not updated in memory. This will reduce the number of failed codewords in the OFEC code. MRBDD fulfills the same purpose as the additional BDD iterations in Algorithm 2 of~\cite{holzbaur_improved_2019}. 
Finally, we define `the half-chunk' to describe the SPR subroutines more easily. Each chunk in the OFEC code can be divided into two half-chunks, the first and second $B$ columns of the relevant chunk, resp. The columns in the first half-chunk of $X(i)$ are coupled with the columns of the second half-chunks of chunks $X(i+3)$ through $X(i+10)$ and vice versa, in the case of $B=16$. 

We check all columns in chunk $X(i)$ and only execute SPR1 in case at least one column has failed to decode or had errors corrected. 
For SPR1, the bits in $X(i)$ are flipped if they meet two conditions: First, the associated column in $X(i)$ either failed to decode or had errors corrected. Second, associated virtual column in chunks $X(i+3)$ through $X(i+N/B+2)$ has failed to decode. After flipping the bits in chunk $X(i)$, the flags of the BDD decoder in the columns of the relevant half-chunks in $X(i)$ through $X(i+N/B)$ are cleared, to prevent the subroutine executing for chunk $X(i+1)$. This is inspired by Alg.~1 of ~\cite{holzbaur_improved_2019}, and helps to avoid generating a new stall pattern as shown in Fig.~5 of ~\cite{holzbaur_improved_2019}.
The other novel subroutine, SPR2, works similarly to SPR1. It flips bits and then clears the decoding result. However, where SPR1 flips bits in $X(i)$, SPR2 flips bits in $X(i+3)$ through $X(i+N/B+2)$. Similar to SPR1, the BDD flags are discarded for chunks $X(i)$ through $X(i+N/B)$ and $X(i+3)$ through $X(i+2(N/B)+2)$.

The SPR3 subroutine is based on the algorithm proposed by Rapp~\cite{rapp_optimized_2024}. 
For the OFEC code, each bit has two associated codewords: one codeword in which it is part of the top $N$ bits, i.e., the main codeword, and another codeword in which it is part of the bottom $N$ (virtual) bits, i.e., the virtual codeword. 
In the original algorithm by Rapp, a bit is flipped if both associated codewords have failed to decode. 
Rather than flipping only upon failures, SPR3 flips a bit when the main codeword fails to decode or has errors corrected, and the virtual codeword fails to decode.

%% file: Figures/OFEC_performance.tikz
\definecolor{Cblue}{RGB}{31 119 180}
\definecolor{Cred}{RGB}{214 39 40}
\definecolor{Corange}{RGB}{255 127 14}
\definecolor{Cgreen}{RGB}{44 160 44}
\definecolor{Cpurple}{RGB}{148 103 198}
\definecolor{Cbrown}{RGB}{140 86 75}
\definecolor{Cpink}{RGB}{227 119 194}
\definecolor{Ccyan}{RGB}{23 190 207}
\definecolor{Cgray}{RGB}{127 127 127}
\definecolor{Cyellow}{RGB}{188 189 34}

\begin{tikzpicture}

\begin{axis}[%
width=1.2\columnwidth,
height=1.2\columnwidth,
at={(0, 0)},
scale only axis,
xmin=13.5,
xmax=15.0,
xtick={13.5, 13.6, ..., 15.1},
x tick label style={rotate=90,anchor=east, /pgf/number format/fixed zerofill, /pgf/number format/precision=1},
ytick={1e-1, 1e-2, 1e-3, 1e-4, 1e-5, 1e-6, 1e-7, 1e-8, 1e-9, 1e-10, 1e-11, 1e-12, 1e-13, 1e-14,  1e-15},
ymode=log,
ymin=9e-16,
ymax=0.1,
yminorticks=true,
axis background/.style={fill=white},
xmajorgrids,
ymajorgrids,
yminorgrids,
xlabel style={font=\color{white!15!black}},
xlabel={SNR [dB]},
ylabel style={font=\color{white!15!black}},
ylabel={post-FEC BER},
legend style={at={(0.99, 0.99)}, anchor=north east, legend cell align=left, align=left, draw=white!15!black}
]
\addplot [color=Cblue, very thick]
  table [x = SNR, y = BER] {Figures/Data/Uncoded.txt};
\addlegendentry{pre-FEC}

\addplot [color=Cred, very thick, mark=*, mark options={solid, Cred , fill=white, scale=1.35}]
  table [x = SNR, y = BER] {Figures/Data/OFEC_no_spr.txt};
\addlegendentry{No SPR}

\addplot [color=Corange, very thick, mark=square*, mark options={solid, Corange , fill=white, scale=1.35}]
  table [x = SNR, y = BER] {Figures/Data/OFEC_spr_min.txt};
\addlegendentry{Rapp's Algorithm \cite{rapp_optimized_2024}}

\addplot [color=Cgreen, very thick, mark=triangle*, mark options={solid, Cgreen , fill=white, scale=1.35}]
  table [x = SNR, y = BER] {Figures/Data/OFEC_spr_B1.txt};
\addlegendentry{Three-stage SPR}

%\addplot [color=Ccyan, very thick, mark=pentagon*, mark options={solid, Ccyan, fill=white, scale=1.35}]
%  table [x = SNR, y = BER] {Figures/Data/OFEC_spr_B1_min2.txt};
%\addlegendentry{Proposed SPR + 2 stages}

%\addplot [color=Cpurple, very thick, mark=diamond*, mark options={solid, Cpurple, fill=white, scale=1.35}]
%  table [x = SNR, y = BER] {Figures/Data/OFEC_spr_B1_min4.txt};
%\addlegendentry{Proposed SPR + 4 stages}

%\addplot [color=Cbrown, very thick, mark=+, mark options={solid, Cbrown , fill=white, scale=1.35}]
%  table [x = SNR, y = BER] {Figures/Data/OFEC_spr_B1_newV4.txt};
%\addlegendentry{Proposed SPR + new stages}

\addplot[mark=none, Black, dashed, samples=2, line width=1.8pt] coordinates {(13.4,1e-10) (15,1e-10)};

\node[anchor=south west] at (14.56, 8e-11) {Sim. by Rapp \cite{rapp_optimized_2024}};

%% New one
\node[anchor=west,align=left, rotate=-40] at (14.11, 5e-12) {Discovered error floor};

\addplot[no markers, white, draw=none]
        table[skip first n=12,
          y={create col/linear regression}
        ] {Figures/Data/OFEC_spr_min.txt};

\addplot[update limits=false, no markers, Corange, thick, dashed, domain={14.05:14.6}]
{exp(x*\pgfplotstableregressiona + \pgfplotstableregressionb)};

\addplot[no markers, white, draw=none]
        table[skip first n=12,
          y={create col/linear regression}
        ] {Figures/Data/OFEC_spr_B1.txt};

\addplot[update limits=false, no markers, Cgreen, thick, dashed, domain={14.05:14.6}]
{exp(x*\pgfplotstableregressiona + \pgfplotstableregressionb)};

%\addplot[mark=none, Black, dashed, samples=2, line width=1.8pt] coordinates {(13.4,1e-10) (15,1e-10)};
%\draw[<->] (14.3) -- ();

%\addplot[no markers, white, draw=none]
%        table[skip first n=11,
%          y={create col/linear regression}
%        ] {Figures/Data/OFEC_spr_set_min.txt};

%\addplot[update limits=false, no markers, Cpurple, dashed, thick, domain={14:14.5}]
%{exp(x*\pgfplotstableregressiona + \pgfplotstableregressionb)};

\end{axis}
\end{tikzpicture}%

%% file: Sections/results.tex
\section{Results and Conclusions}
%Figure~\ref{fig:results} shows the results of the novel SPR approach for 16-QAM.
Fig.~\ref{fig:results} shows the performance of the discussed algorithms, simulated using 16-QAM over an AWGN channel. 
The simulations were done using C++/CUDA code at a 1~Gbps simulation throughput per NVIDIA RTX 4090 GPU. 
An error floor can be seen at a post-FEC BER of $10^{-9}$ without any SPR. 
Algorithm by Rapp~\cite{rapp_optimized_2024} removes this error floor.
However, a second error floor is observed at $10^{-12}$. 
Our novel three-stage SPR algorithm removes this second error floor.
However, we observe a new error floor at $10^{-13}$, i.e., obtain an order of magnitude improvement in the flooring region. 
We estimate via extrapolation that HD SOTA~\cite{rapp_optimized_2024} achieves $10^{-15}$ at 14.5~dB SNR. 
Our algorithm is estimated to achieve $10^{-15}$ at $14.35$~dB, indicating a $0.15$~dB gain.
These results show that error-floor-free HD-only decoding of OFEC is feasible with more complex SPR algorithms.

%% file: Figures/SPR_pipeline.tikz
%\definecolor{Cblue}{RGB}{31 119 180}
%\definecolor{Cred}{RGB}{214 39 40}
%\definecolor{Corange}{RGB}{255 127 14}
\definecolor{Cgreen}{RGB}{44 160 44}
%\definecolor{Cpurple}{RGB}{148 103 198}
\definecolor{Cbrown}{RGB}{140 86 75}
%\definecolor{Cpink}{RGB}{227 119 194}
%\definecolor{Ccyan}{RGB}{23 190 207}
%\definecolor{Cgray}{RGB}{127 127 127}
%\definecolor{Cyellow}{RGB}{188 189 34}
\begin{tikzpicture}

    \node[draw, minimum width=1.4cm, minimum height=0.7cm] (BDD) at (0, 0) {BDD};
    \node[draw, minimum width=1.4cm, minimum height=0.7cm] (SP1) at (2, 0) {SPR1};
    \node[draw, minimum width=2cm, minimum height=0.7cm] (SP2) at (4.78, 0) {SPR2};
    \node[draw, minimum width=1.4cm, minimum height=0.7cm] (SP3) at (7.5, 0) {SPR3};

    %\node[draw, minimum width=1.4cm, minimum height=0.7cm] (SP4) at (7.5, -1) {SPR2};
    %\node[draw, minimum width=1.4cm, minimum height=0.7cm] (SP5) at (5, -1) {SPR3a};
    %\node[draw, minimum width=1.4cm, minimum height=0.7cm] (SP6) at (2.5, -1) {SPR3b};
    %\node[draw, minimum width=1.4cm, minimum height=0.7cm] (SP7) at (0, -1) {SPR3c};

    \node[draw, rotate=90, anchor=north, minimum width=0.7cm, inner sep=1.3pt] (SP1BDD) at (SP1.east) {\resizebox{.6cm}{!}{BDD}};
    \node[draw, rotate=90, anchor=north, minimum width=0.7cm, inner sep=1.3pt] (SP2BDD) at (SP2.east) {\resizebox{.6cm}{!}{BDD}};
    \node[draw, rotate=90, anchor=north, minimum width=0.7cm, inner sep=1.3pt] (SP3BDD) at (SP3.east) {\resizebox{.6cm}{!}{BDD}};

    \node[draw, rotate=90, anchor=south, minimum width=0.7cm, inner sep=1.3pt] (SP1MRBDD) at (SP1.west) {\resizebox{.6cm}{!}{MRBDD}};
    \node[draw, rotate=90, anchor=south, minimum width=0.7cm, inner sep=1.3pt] (SP2MRBDD) at (SP2.west) {\resizebox{.6cm}{!}{MRBDD}};
    \node[draw, rotate=90, anchor=south, minimum width=0.7cm, inner sep=1.3pt] (SP3MRBDD) at (SP3.west) {\resizebox{.6cm}{!}{MRBDD}};

    %\node[draw, rotate=90, anchor=south, minimum width=0.7cm, inner sep=1.3pt] (SP4BDD) at (SP4.west) {\resizebox{.6cm}{!}{BDD}};
    %\node[draw, rotate=90, anchor=south, minimum width=0.7cm, inner sep=1.3pt] (SP5BDD) at (SP5.west) {\resizebox{.6cm}{!}{BDD}};
    %\node[draw, rotate=90, anchor=south, minimum width=0.7cm, inner sep=1.3pt] (SP6BDD) at (SP6.west) {\resizebox{.6cm}{!}{BDD}};
    %\node[draw, rotate=90, anchor=south, minimum width=0.7cm, inner sep=1.3pt] (SP7BDD) at (SP7.west) {\resizebox{.6cm}{!}{BDD}};
    
    %\node[draw, rotate=90, anchor=north, minimum width=0.5cm, inner sep=1.3pt] (SP4MRBDD) at (SP4.east) {\resizebox{.4cm}{!}{MRBDD}};

    \draw[-Stealth, line width=0.7pt] (BDD) -- (SP1MRBDD);
    \draw[-Stealth, line width=0.7pt] (SP1BDD) -- (SP2MRBDD);
    \draw[-Stealth, line width=0.7pt] (SP2BDD) -- (SP3MRBDD);
    %\draw[->, line width=0.7pt] (SP3BDD) --++ (-.5, 0) --++ (0, -.5) -- (SP4 |- 53, -0.5) --++ (1, 0) --++ (0, -.5) -- (SP4);
    
    %\draw[->, line width=0.7pt] (SP4BDD) -- (SP5);
    %\draw[->, line width=0.7pt] (SP5BDD) -- (SP6);
    %\draw[->, line width=0.7pt] (SP6BDD) -- (SP7);

    %\draw[line width=1.6pt, color=Cbrown] (-1.6, 0.9) -- (8.8, 0.9) -- (8.8, -1.5) -- (-1.6, -1.5) -- (-1.6, 0.9);
    %\node[anchor = south west] at (-1.6, 0.82){\textcolor{Cbrown}{Algorithm 2}};

    %\draw[line width=1.6pt, color=Cgreen] (-1.5, 0.8) -- (8.7, 0.8) -- (8.7, -0.45) -- (-1.5, -0.45) -- (-1.5, 0.8);
    %\node[anchor = north west] at (-1.5, 0.82){\textcolor{Cgreen}{Algorithm 1}};
    \node[anchor = center] at (2, -0.55) {\scalebox{.75}{Flip in $X(i)$}};
    \node[anchor = center] at (2, -0.9) {\scalebox{.75}{Remove flags}};

    \node[anchor = center] at (4.78, -0.55) {\scalebox{.75}{Flip in $X(i+3)$ to $X(i+10)$}};
    \node[anchor = center] at (4.78, -0.9) {\scalebox{.75}{Remove flags}};

    \node[anchor = center] at (7.5, -0.55) {\scalebox{.75}{Flip in $X(i)$}};
    \node[anchor = center] at (7.5, -0.9) {\scalebox{.75}{Don't remove flags}};
    
\end{tikzpicture}

%% file: references.bib
@article{pyndiah_near-optimum_1998,
	title = {Near-optimum decoding of product codes: block turbo codes},
	volume = {46},
	issn = {00906778},
	doi = {10.1109/26.705396},
	shorttitle = {Near-optimum decoding of product codes},
	pages = {1003--1010},
	number = {8},
	journaltitle = {{IEEE} Transactions on Communications},
	shortjournal = {{IEEE} Trans. Commun.},
	author = {Pyndiah, R.M.},
	date = {1998-08},
    year = {1998},
    journal = {IEE Transactions on Communications},
	langid = {english},
}

@article{zokaei_memory_2022,
	title = {Memory Optimized Hardware Implementation of Open {FEC} Encoder},
	volume = {30},
	issn = {1063-8210, 1557-9999},
	doi = {10.1109/TVLSI.2022.3180554},
	pages = {1548--1552},
	number = {10},
    journal={IEEE Transactions on Very Large Scale Integration (VLSI) Systems}, 
	shortjournal = {{IEEE} Trans. {VLSI} Syst.},
	author = {Zokaei, Abolfazl and Truhachev, Dmitri and El-Sankary, Kamal},
	date = {2022-10},
    year = {2022},
	langid = {english},
}

@article{bose_class_1960,
	title = {On a class of error correcting binary group codes},
	volume = {3},
	issn = {00199958}, 
	doi = {10.1016/S0019-9958(60)90287-4},
	pages = {68--79},
	number = {1},
	journaltitle = {Information and Control},
	shortjournal = {Information and Control},
    journal = {Information and Control},
    year = {1960},
	author = {Bose, R.C. and Ray-Chaudhuri, D.K.},
	date = {1960-03},
	langid = {english},
}

@article{chase_class_1972,
	title = {Class of algorithms for decoding block codes with channel measurement information},
	volume = {18},
	issn = {0018-9448},
	doi = {10.1109/TIT.1972.1054746},
	pages = {170--182},
	number = {1},
	journaltitle = {{IEEE} Transactions on Information Theory},
    journal={IEEE Transactions on Information Theory}, 
	shortjournal = {{IEEE} Trans. Inform. Theory},
	author = {Chase, D.},
	date = {1972-01},
	langid = {english},
    year = {1972},
}

@article{wang_real-time_2023,
	title = {Real-Time {FPGA} Investigation of Potential {FEC} Schemes for 800{G}-{ZR}/{ZR}+ Forward Error Correction},
	volume = {41},
	issn = {0733-8724, 1558-2213},
	doi = {10.1109/JLT.2022.3218957},
	pages = {926--933},
	number = {3},
	journaltitle = {Journal of Lightwave Technology},
	shortjournal = {J. Lightwave Technol.},
	author = {Wang, Weiming and Long, Zhijun and Qian, Weifeng and Tao, Kai and Wei, Zitao and Zhang, Shihua and Feng, Zhenhua and Xia, Yan and Chen, Yong},
	date = {2023-02-01},
	langid = {english},
    year = {2023},
    journal = {Journal of Lightwave Technology},
}

@article{holzbaur_improved_2019,
	title = {Improved decoding and error floor analysis of staircase codes},
	volume = {87},
	issn = {0925-1022, 1573-7586},
	doi = {10.1007/s10623-018-0587-x},
	pages = {647--664},
	number = {2},
	journaltitle = {Designs, Codes and Cryptography},
    journal = {Designs, Codes and Cryptography},
	shortjournal = {Des. Codes Cryptogr.},
	author = {Holzbaur, Lukas and Bartz, Hannes and Wachter-Zeh, Antonia},
	date = {2019-03},
    year = {2019},
	langid = {english},
}

@standard{open_roadm_2022,
    title = {{Open ROADM MSA 6.0 W-Port Digital Specification (400G-800G)}},
    author = {Sluyski, Mike A.},
    date = {2022-01-11},
    institution = {OpenROADM},
}

@standard{open_zr_MSA,
    title = {{Open ZR+ MSA Technical Specification}},
    author = {Srivastava, Atul and  Williams, Tom and Zhang, Bo},
    date = {2023-09-12},
    institution = {OpenZR+ MSA Group},
}

@article{hocquenghem_codes_1959,
    title = {Codes Correcteurs d’Erreurs},
    date = {1959},
    author = {A. Hocquenghem},
    journaltitle = {Chiffres},
    journal = {Chiffres},
    year = {1959},
}

@conference{rapp_optimized_2024,
  title={Optimized Soft-Aided Decoding of OFEC and Staircase Codes}, 
  author={Lukas Rapp and Sisi Miao and Laurent Schmalen},
  year={2024},
  eprint={2404.19532},
  archivePrefix={arXiv},
  primaryClass={cs.IT},
  booktitle={50th European Conference on Optical Communications (ECOC 2024)}, 
  volume={2024},
  number={},
}

@ARTICLE{sukmadji_zipper_2022,
  author={Sukmadji, Alvin Y. and Martínez-Peñas, Umberto and Kschischang, Frank R.},
  journal={Journal of Lightwave Technology}, 
  title={Zipper Codes}, 
  year={2022},
  volume={40},
  number={19},
  pages={6397-6407},
  doi={10.1109/JLT.2022.3193635}
}

@ARTICLE{hager_approaching_2018,
  author={Häger, Christian and Pfister, Henry D.},
  journal={IEEE Transactions on Communications}, 
  title={Approaching Miscorrection-Free Performance of Product Codes With Anchor Decoding}, 
  year={2018},
  volume={66},
  number={7},
  pages={2797-2808},
  doi={10.1109/TCOMM.2018.2816073}
}

@ARTICLE{nagarajan_low_2021,
  author={Nagarajan, Radhakrishnan and Lyubomirsky, Ilya and Agazzi, Oscar},
  journal={Journal of Lightwave Technology}, 
  title={Low Power {DSP}-Based Transceivers for Data Center Optical Fiber Communications (Invited Tutorial)}, 
  year={2021},
  volume={39},
  number={16},
  pages={5221-5231},
  doi={10.1109/JLT.2021.3089901}
}

@ARTICLE{zhou_beyond_2020,
  author={Zhou, Xiang and Urata, Ryohei and Liu, Hong},
  journal={Journal of Lightwave Technology}, 
  title={{Beyond 1 Tb/s Intra-Data Center Interconnect Technology: IM-DD or Coherent?}}, 
  year={2020},
  volume={38},
  number={2},
  pages={475-484},
  doi={10.1109/JLT.2019.2956779}
}

@book{barroso_datacenter_2018,
    author={Barroso, Luiz André and Hölzle, Urs and Ranganathan, Parthasarathy},
    title={{The Datacenter as a Computer: An Introduction to the Design of Warehouse-Scale Machines}},
    year={2018},
    publisher={Morgan \& Claypool},
    address={San Rafael, CA, USA},
}
